\documentclass[conference]{IEEEtran}
\usepackage{booktabs}
\usepackage{balance}
\usepackage{graphics}
\usepackage[T1]{fontenc}
\usepackage{multirow}
\usepackage{enumitem}
\usepackage{graphicx}
\usepackage{mdwlist}
\usepackage{tabularx}
\usepackage{color}
\usepackage{soul}
\usepackage{ifthen}
\usepackage{comment}
\usepackage{bm}

\newcommand{\name}{CarePre~}

\begin{document}

\linespread{0.9}
\title{CarePre: An Intelligent Clinical Decision \\ Assistance System}
\author{\small{Zhuochen Jin$^{1}$, Jingshun Yang$^{1}$, Shuyuan Cui$^{1}$, David Gotz$^{2}$, Jimeng Sun$^{3}$, Nan Cao$^{1*}$}\\
1. Intelligent Big Data Visualization Lab, Tongji University, China\\
$^*$ \texttt{nan.cao@gmail.com} \\
2. University of North Carolina at Chapel Hill, USA\\
3. Georgia Institute of Technology, USA\\
}

\maketitle

\begin{abstract}
Clinical decision support systems (CDSS) are widely used to assist with medical decision making.  However, CDSS typically require manually curated rules and other data which are difficult to maintain and keep up-to-date. Recent systems leverage advanced deep learning techniques and electronic health records (EHR) to provide a more timely and precise results. Many of these techniques have been developed with a common focus on predicting upcoming medical events.  However, while the prediction results from these approaches are promising, their value is limited by their lack of interpretability. To address this challenge, we introduce \name, an intelligent clinical decision assistance system. The system extends a state-of-the-art deep learning model to predict upcoming diagnosis events for a focal patient based on his/her historical medical records. The system includes an interactive framework together with intuitive visualizations designed to support diagnosis, treatment outcome analysis, and the interpretation of the analysis results. We demonstrate the effectiveness and usefulness of \name system by reporting results from a quantities evaluation of the prediction algorithm and a case study and three interviews with senior physicians.
\end{abstract}
\section{Introduction}
\label{sec:intro}



Medical decision making is fraught with uncertainty. It involves not only deciding what disease a patient may have, but also which treatments to choose from a set of possible alternatives~\cite{musen2014clinical}. Motivated by these challenges, clinical decision support systems (CDSS) have gained increasing usage in recent years. CDSS are 
computer-based systems which integrate into the clinical workflow to help physicians determine which questions to ask, which tests to order, and which procedures to perform \cite{rais2011operations, salehipour2012exact}. However, typical CDSS require manually curated knowledge bases that are difficult to maintain and keep up-to-date, thus limiting their usage in real world clinical scenarios~\cite{marakas2003decision}.

The rapid development of machine learning techniques and the increasing availability of electronic health records (EHR) has stimulated great interest in harnessing EHR data to  help drive CDSS. It is widely believed that high quality EHR data in the context of CDSS has the potential to reduce errors and provide more precise results~\cite{blumenthal2010meaningful,carter2001electronic,carter2007design,kuperman1999clinical}. To this end, many techniques have been developed to extract meaningful insights from EHR data with a common focus on prediction of upcoming medical events (e.g., a diagnosis or treatment)~\cite{crosson2005implementing,grant2008practice}. In particular, a series of deep learning-based prediction models~\cite{choi2016doctor,choi2016retain,xiao2017joint} have successfully demonstrated that high accuracy predictions are possible.  However, the utility of these methods is greatly limited by their lack of interpretability.  The ideal intelligent medical event prediction system must provide results that are both accurate and interpretable through a user-friendly interface.

However, achieving both accuracy and interpretability is challenging as they are often achieved via contradictory design decisions.  The highest accuracy prediction is often obtained when using more complex prediction methods, whereas simpler models with lower accuracy are often more interpretable~\cite{breiman2001statistical}.
Attempts have been made to improve the interpretability of more complex prediction models~\cite{choi2016retain,xiao2017joint}. However, these approaches are still too complex for users with little or no technical training, such as medical doctors.

To address the above issues, we introduce \name, an intelligent clinical decision assistance system. \name predicts the risks of a patient being diagnosed in the future with certain diseases based on his/her historical electronic health records. The system extends a state-of-the-art deep prediction model that is specifically designed for predicting medical events, and employs intuitive visualization techniques to help interpret the prediction results without reducing the complexity of the underlying model. In particular, \name supports interpretation by (1) framing the prediction results in the context of a group of similar patients, and (2) analyzing the factors that influence the prediction results to help physicians make a more informed clinical decision. The contributions of the paper include:
\begin{itemize}
    \item \textbf{System Design.} We introduce a comprehensive clinical decision assistance system for predicting a patient's risk of future diagnosis for certain diseases, and estimating the outcome of different treatments based on the patient's electronic health records. The system design is guided by results from a pilot study with two senior physicians.
    \item \textbf{Exploratory Analysis.} We propose an interactive framework that supports detailed exploration for both (1) interpretation of prediction results in the context of historical and similar medical records, and (2) analysis of potential treatment outcomes.
    \item \textbf{Evaluation.} We evaluate the system via a quantitative evaluation of the algorithm, a case study with a senior physician, and three interviews with three case studies using real-world medical data with three senior physicians. We describe the case studies and results from interviews with each physician. These results provide evidence regarding the usefulness of the system.
\end{itemize}

\section{Related Works}
\label{sec:related}
In this section, we provide an overview of previous research that is most relevant to our work including: (1) clinical decision support systems (CDSS), (2) prediction models in medicine, and (3) visualization of electronic health records.

\subsection{Clinical Decision Support Systems}
Existing clinical decision support systems (CDSS) can be primarily summarized into two major types: Knowledge-based and Non-knowledge-based \cite{berner2007overview}.
Knowledge-based systems, which are are most commonly used, typically organize knowledge about diseases and the associations of symptoms in the form of if-then rules. For example, Dayan et al.~\cite{dayan2017use} introduced the traumatic brain injury (TBI) prediction rules in a CDSS to foresee risks of TBI. Laleci et al.~\cite{laleci2018personalised} utilized a guideline-based CDSS to help manage the personal care plans of elders. Rodriguez et al.~\cite{rodriguez2018send} introduced a ``send \& hold'' system, utilizing clinical decision support rules to reduce the avoidable vitamin testing.

Non-knowledge-based systems are usually developed based on machine learning techniques that can automatically learn the associations between symptoms and diseases from electronic health record (EHR) data~\cite{berner2007overview}. It has been shown that EHR data not only helps improve the precision of analysis results~\cite{blumenthal2010meaningful,emani2017physician}, but also greatly improves the robustness of a CDSS due to the availability of rich and diverse EHR data gathered during the daily clinical encounters~\cite{choi2016doctor,wu2017omic}. When compared to knowledge-based systems, these systems greatly reduce the human efforts required to manually build and update a large knowledge database~\cite{marakas2003decision}. However, these systems typically suffer from a lack of interpretablity of the analysis results~\cite{marakas2003decision}, and a lack of user-friendly interfaces to facilitate efficient results inspection~\cite{berner1994performance,carter2007design}. \name leverages the advantages of machine learning techniques and electronic health records, while also providing a comprehensive visualization-based design to support result inspection and interpretation.

\subsection{Prediction Models in Medicine}
Prediction models have played an increasingly important role in the medical domain, for both diagnosis and prognosis~\cite{steyerberg2008clinical}. Recent research has often focused on leveraging deep learning techniques to make predictions more accurate and precise~\cite{Yao2018}. These techniques have been used to support public health analysis~\cite{chainani2018disease, zhao2015simnest,zou2016infectious}, medical research~\cite{unterthiner2015toxicity,danaee2017deep,yousefi2017predicting}, and clinical practice~\cite{acharya2017application,lawhern2018eegnet,jagannatha2016bidirectional}. Some deep learning techniques have been developed to assess risk for specific conditions, such as the diagnosis of heart disease~\cite{acharya2017application,rajpurkar2017cardiologist, yan2015restricted}, cancer~\cite{chaudhary2017deep,danaee2017deep,yousefi2017predicting}, and mental health~\cite{benton2017multi, hammerla2015pd,ravi2016deep}.  

Most relevant to our work, other research has focused on predicting upcoming medical events (e.g., a future diagnosis or treatment) based on electronic health records (EHR).  Examples in this area include Jagannatha et al.~\cite{jagannatha2016bidirectional}, who used EHR data to train a bidirectional recurrent neural network (RNN) for medication and disease prediction. Choi et al.~\cite{choi2016doctor} developed Doctor AI, a generic RNN model that use historical EHR to predict the clinical events as well as the time to the next visit. Following this work, Choi et al.~\cite{choi2016retain} further introduced Retain, a state-of-the-art, high-accuracy prediction model that was specifically designed to predicting `signal' events (i.e, heart failure) based on EHR data. Our system extends this model to predict multiple events, as motivated by our design requirements. 

Interpreting results from prediction models is a recognized challenge, and it is especially difficult for models that leverage deep learning. Existing interpretation techniques can largely categorized into two categories: (1) global model analysis, which employs visualization techniques to represent the internal structure of a deep learning model~\cite{liu2017towards,strobelt2018lstmvis,liu2018analyzing}, and (2) instance-based analysis, which monitors changes to results in response to changes in model input~\cite{krause2016interacting,nguyen2015deep}. \name adopts the instance-based analysis approach via a set of interactive visualization designs that allow users to adjust/delete/add medical events within a patients' historical medical records explore their impact on the prediction result.

\subsection{Visual Analysis of Electronic Health Records}
Many visual analysis systems have been developed for representing and analyzing electronic health records. Most of these summarize a large set of EHR data into a flow-based representation that reveals the frequent patterns of medical event sequences~\cite{monroe2013temporal,perer2015mining} and the outcomes yielded by different care plans~\cite{wongsuphasawat2011outflow,perer2013data}. However, these techniques are typically challenged by event sequences of varied length that contain large numbers of event types.
These real-world properties of medical data can often lead to cluttered and less meaningful visualizations when sequences vary dramatically. To overcome this limitation, Gotz et al.~\cite{gotz2014decisionflow} introduced DecisionFlow, in which sequences with different length and large numbers of even types are visualized based on several key events.   This hides the complexity introduced by other non-key event types. Guo et al.~\cite{guo2018visual} introduced ET$^2$, in which the sequences are aligned based on dynamic time wrapping and segmented into stages shown with more details to help illustrate the progression of a disease in context of a care plan. Du et al.~\cite{du2016eventaction,du2017finding} introduced visual analysis systems to predict upcoming events or recommend the next procedure by summarizing a set of similar event sequences without using any prediction model, thus producing results with limited accuracy. \name leverages many of the advances contributed by these visualization techniques, and supports multiple visualization-based views to help users exploration and interpret prediction results.

\section{Pilot Study}
\label{sec:system}

\begin{figure*}[!htb]
    \centering
    \includegraphics[width=\linewidth]{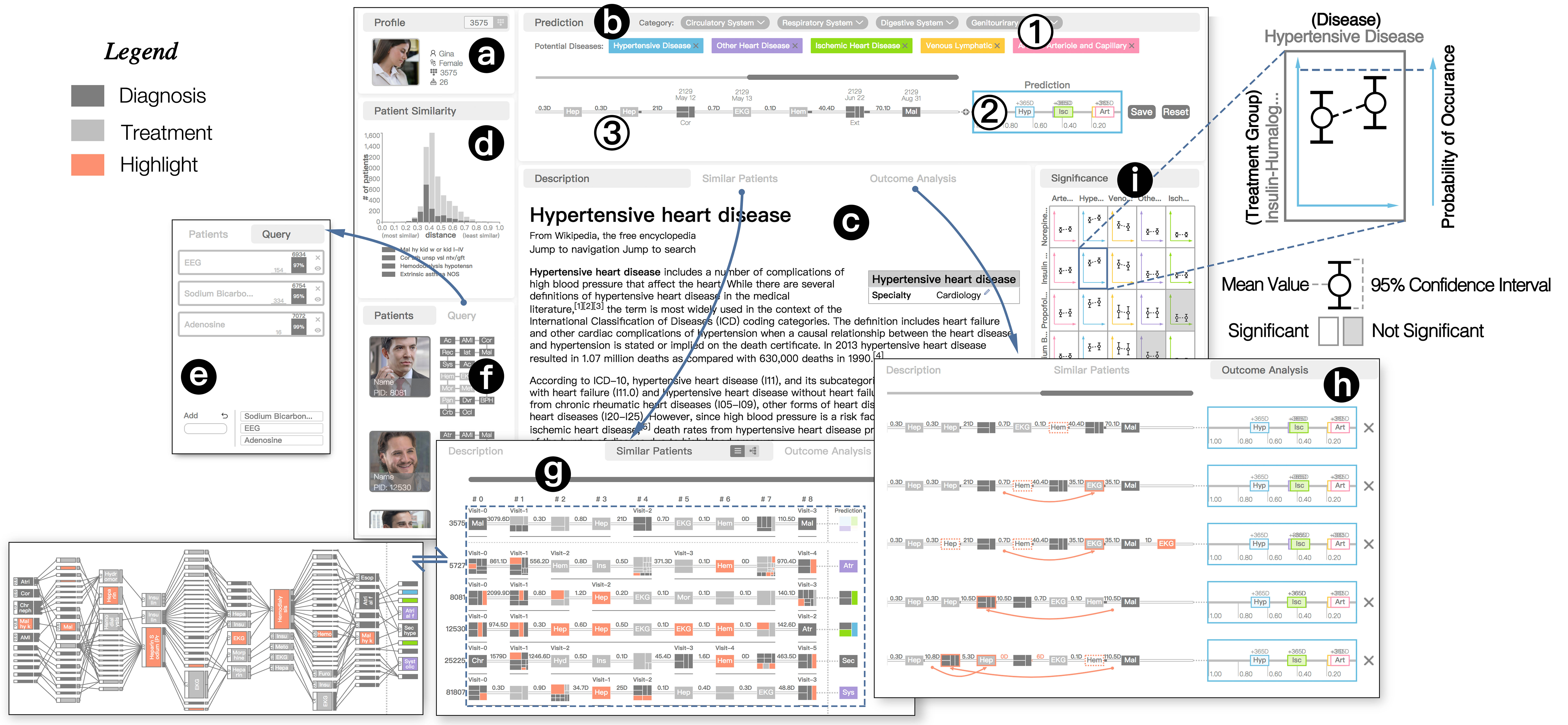}
    \caption{The \name system contains nine interactively coordinated views, including (a) a profile view showing the personal information of a patient; (b) a prediction view illustrating the prediction results as well as the historical medical records of the patient; (c) a description view providing the detailed description of a disease selected from the prediction view, (d) a patient similarity view measuring the similarity between the focal patient and the archived patients; (e) a query view supporting a key-event-based query capability to select specific patients; (f) a patient list showing similar patients retrieved from (d) or (e); (g) a similar patients view comparing the prediction results to the outcomes of similar patients; (h) an outcome analysis view allowing the examination of the outcomes of different treatment plans; and (i) a significance view showing the influence of treatments on the risks of diseases.}
    \label{fig:interface} 
    \vspace{-0.4cm}
\end{figure*}

Our pilot study followed a multi-session design, and involved two senior physicians with over 15 year's clinical experience in two hospitals in a major city in China. The goal of the pilot study was to determine detailed requirements to guide the subsequent system design.

\textbf{Session~1: Initial Requirements.} Interviews were performed with each of the two senior physicians, during which we discussed the challenges that they encountered during their daily work. Both physicians, although experienced, expressed a fear of making mistakes.  They see many patients each day, and typically don't have enough time to fully review a patient's medical history.  This means that diagnosis and treatment decisions are often made based on the patient's \emph{current} symptoms and lab test results. They expressed the desire for a system that could: (1) automatically provide relevant diagnosis information based on a patient's medical history (i.e., diagnosis-supporting requirements); and (2) assist in estimating potential outcomes if the doctor were to administer a particular treatment (prognosis-supporting requirements). 

\textbf{Session~2: Prototyping and Refinement.} Based on results of Session~1, an interactive design prototype was developed using figma\footnote{\url{https://www.figma.com/}} by a professional designer (a co-author of this article).  The prototype was demonstrated to the two doctors to solicit feedback, leading to the following more detailed requirements:

\begin{itemize}
    \item [\textbf{R1}] The system should be able to automatically assess a patient's historical medical record to predict the risks of a set of potential diseases identified by the physicians.

    \item [\textbf{R2}] The system should be able to illustrate the prediction results within the context of the patient's historical medial record to facilitate data exploration and result interpretation.
    
    \item [\textbf{R3}] The system should support easy comparison between the focal patient and other patients with similar historical medical records.  This would help a physician further verify and interpret the prediction results.
    
    \item [\textbf{R4}] The system should be able to identify and communicate the key factors that would increase or decrease a patient's risk.  Furthermore, the system should allow physicians explore changes to possible treatment plans and understand the impact of those changes on the predicted outcome.
\end{itemize}

The entire prototyping stage took place over two months during which regular meetings with domain experts were held.  The prototype was iteratively refined to incorporate  clinician comments and new requirements. This process resulted in a series of eight different design versions, culminating in the final design described in the next section.
\section{\name System}
Following the aforementioned requirements, we designed the \name system.  This section provides an overview of the system design and its key algorithms.

\subsection{System Overview}
\name is an intelligent system designed to assist physicians or other health professionals when making decisions related to diagnosis and prognosis. The key functionalities of \name are (1) prediction of a patient's risk of being diagnosed with certain disease, and (2) estimation of the most influential treatments, as determined based on a patient's historical electronic health records (EHR). In particular, the system coverts raw EHR data for a large number of patients into sequences of medical events. Based on those sequences, the system predicts for a focal patient the future occurrence probabilities of several given diagnosis events.

Fig.~\ref{fig:interface} illustrates the \name user interface.  It consists of ten views, many of which utilize data visualization techniques to facilitate an intuitive data representation and interpretation. These views can be categorized into three classes based on their functionality: (1)~diagnosis-supporting views (Fig.~\ref{fig:interface}(a,b,c)), (2)~similar patients retrieval and comparison views (Fig.~\ref{fig:interface}(d-g)), (3)~treatment outcome analysis views (Fig.~\ref{fig:interface}(b,h,i)). 

\begin{figure}[!tb]
    \centering
    \includegraphics[width=\columnwidth]{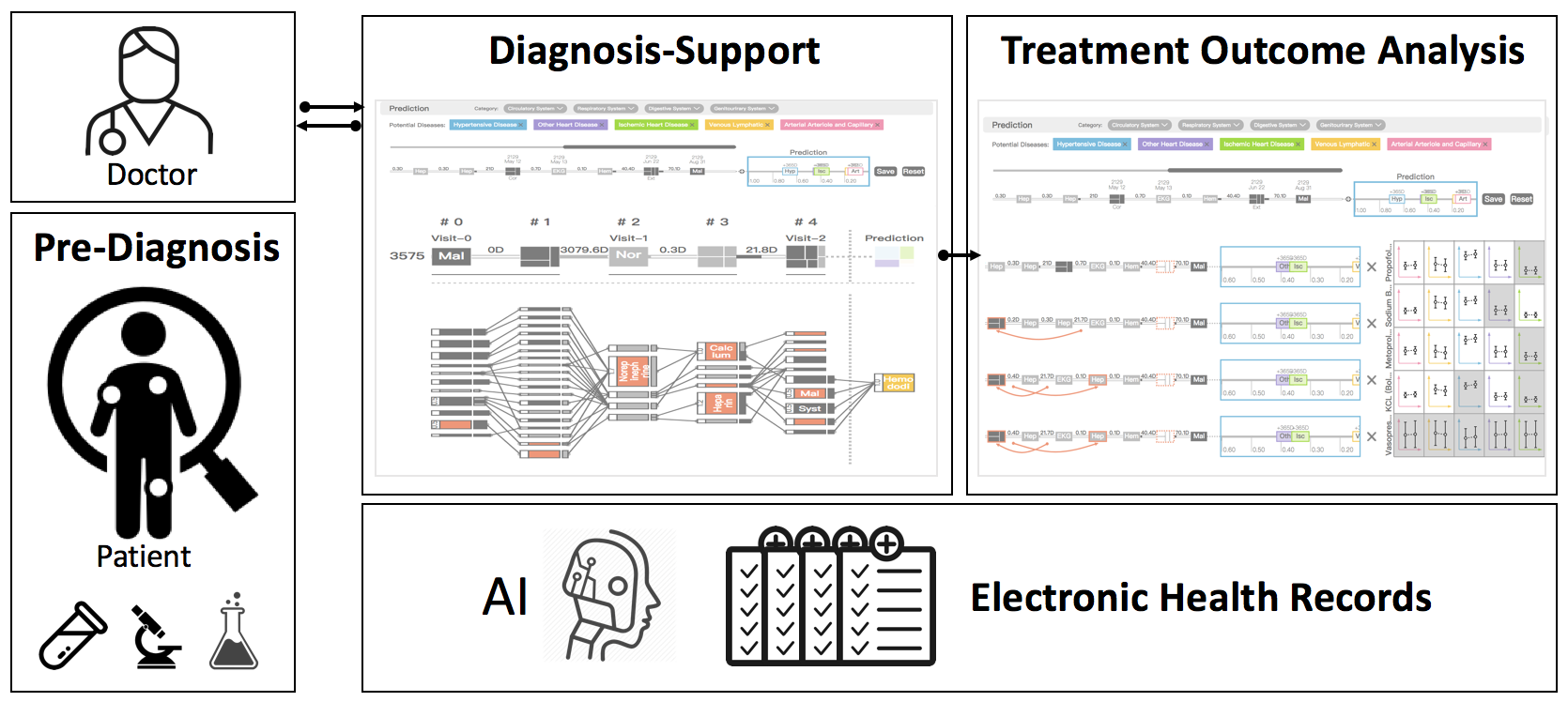}
    \caption{The interaction pipeline of \name system consists of three steps, including: (1) a pre-diagnosis step in which physicians initially diagnosis a focal patient according to his/her symptoms and lab tests; (2) a diagnosis-support step in which the system automatically estimates the risk of each the potential diseases determined in the previous step, and in which physicians can verify the results by comparison to the medical records of a set of similar patients; and (3) the treatment outcome analysis step in which physicians can compare and evaluate the expected outcomes of different treatment plans.}
    \label{fig:interaction} 
    \vspace{-0.5cm}
\end{figure}

These views support the system's interaction pipeline as shown in Figure~\ref{fig:interaction}. The pipeline includes three main steps, beginning with a physician making an initial diagnosis using his/her own knowledge and experiences, based on a focal patient's current symptoms and lab tests. The potential diagnoses from this stage or entered into the system (Fig.~\ref{fig:interface}(b-1)), which automatically estimates the patient's risks in terms of being diagnosed in the future with the diseases given his/her historical medical records (Fig.~\ref{fig:interface}(b-2)). The doctor can explore the details of the historical medical records (Fig.~\ref{fig:interface}(b-3)), and issue a query to fetch a set of similar patients to help contextualize and interpret the prediction results (Fig.~\ref{fig:interface}(d,f,g)).  Third, the doctor can examine alternative treatment plans by examining and comparing the expected outcomes of each as estimated by the system (Fig.~\ref{fig:interface}(h,i)).

\subsection{Diagnosis Support}

\name system assists a typical diagnosis procedure by predicting the next medical event given an event sequence representing a patient's medical record (\textbf{R1}). More specifically, the system, the system predicts the next diagnosis (i.e., the potential diseases a patient may have) based on the patient's previous diagnoses and treatments. The prediction results and the patient's historical medical data are illustrated in an interactive timeline visualization to facilitate data and result exploration (\textbf{R2}). 

\begin{figure}[!tb]
    \centering
    \includegraphics[width=\columnwidth]{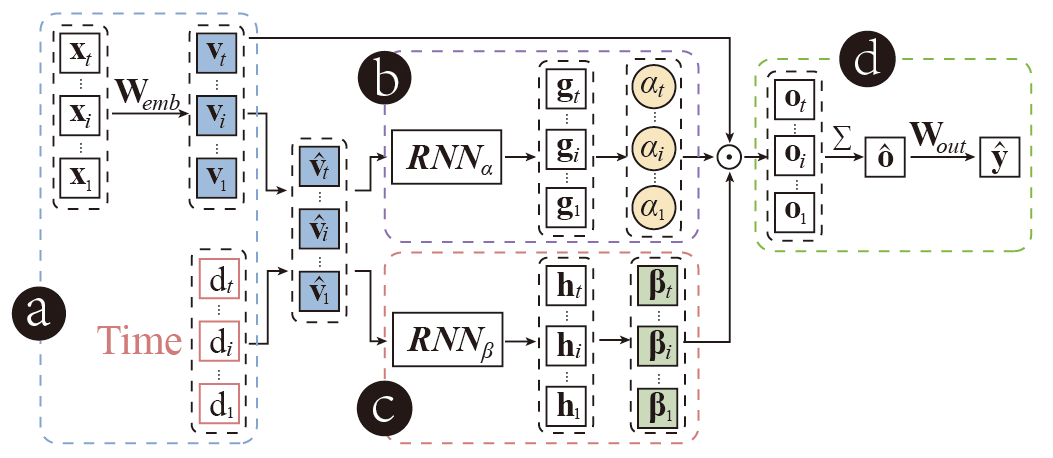}
    \caption{The structure of Retain: Taken a sequence $x_{1},\cdots,x_{t}$ as input, the model can predict the distribution of possible diagnosis in three steps: (a) the embedding step, (b,c) the attention steps, (d) the final prediction step.}
    \label{fig:retain} 
    \vspace{-0.4cm}
\end{figure}

\subsubsection{Prediction Model} To predict the next diagnosis, we utilize a deep learning model with two recurrent neural networks~\cite{hochreiter1997long} to predict the likelihood of occurrence for a set of potential diseases selected by physicians based on a patient's historical medical record. The model extends the design used in Retain~\cite{choi2016retain} to predict multiple medical events at the same time. Our model was trained using a subset of the MIMIC dataset~\cite{johnson2016mimic}, which contains the electronic health records of 46,521 patients. Prior to training, the data were cleaned by removing the low frequency and irrelevant event types. 

Fig.~\ref{fig:retain} illustrates the structure of the model.  The model predicts subsequent medical events based on an input event sequence $[x_1,..., x_t]$, where $x_i$ is a multi-hot vector that captures the occurrences of events at each time point. Given this input, an embedding layer is used to project each of the input events into a latent feature vector $v_i$ (Fig.~\ref{fig:retain}(a)). After that, $v_i$ is further concatenated with $d_i$, the duration between the i-$th$ event in the sequence and the prediction time, which is denoted as $\hat{v}_i = [v_i,d_i]$.  This combined vector is the input for two recurrent neural networks (RNNs) as shown in Fig.~\ref{fig:retain}(b,c). 

The first network, $RNN_\alpha$ (Fig.~\ref{fig:retain}(b)), takes the information of all events at each time points into consideration to ensure a high accuracy prediction result. The outputs of the model, i.e., $(\alpha_1,...,\alpha_t)$ are weights that indicate the accumulated influence on the prediction results at each time point.  

The second network, $RNN_\beta$ (Fig.~\ref{fig:retain}(c)), estimates the influence of each individual event at each time point in time on the prediction results.  These estimates facilitate interpretation of the prediction results. The output, $(\beta_1,...,\beta_t)$, are vectors at different time points with each field in a vector representing the influence of an individual event on the prediction results. A positive / negative field value corresponds to an event that is associated with an increase / decrease in the occurrence probability of the predicted event, respectively. 

The results from the overall model are calculated using a softmax layer, which predicts the occurrence probability of each event as follows:
\[
\hat{\textbf{y}}_{t} = softmax(\textbf{W}_{out}\hat{\textbf{o}}_{t}+\textbf{e}_{out})
\]
where $\textbf{W}_{out}$ and $\textbf{e}_{out}$ are the parameters to be learned in the softmax-layer; $\hat{\textbf{o}}_{t}$ is the context vector at time point $t$, which we define as a combination of the previous outputs: 
\[
\hat{\textbf{o}}_{t} = \sum_{i=1}^{t}\alpha_{i}\bm{\beta}_{i}\odot\textbf{v}_{i}
\]
where $\odot$ is the element-wise multiplication operator.

The model is trained based on the following loss function:
\[
L = -\frac{1}{N}\sum_{k=1}^{N}\frac{1}{T^{(k)}}\sum_{t=1}^{T^{(k)}}(\textbf{b}_{w}\textbf{y}_{t}^{T}log(\hat{\textbf{y}}_{t})+(1-\textbf{y}_{t})^{T}log(1-\hat{\textbf{y}}_{t}))
\]
where $N$ is the number of samples, $T^{(k)}$ is the length of the sequence in each sample, $y_{t}$ is the ground truth, and $\hat{y}_{t}$ represents the prediction results. $b_{w}$ is a vector that is included within the loss function to address the presence of highly skewed training data. Each field in $b_{w}$ is calculated as $1/log(n)$ where $n$ indicates the number of occurrences of an event within the training samples. $b_w$ helps overcome skewed distributions within the training samples by reducing the marginal importance of additional event occurrences for high frequency events. Finally, we estimate the influence of a historical event $s$ occurring at timestamp $t$ to the prediction results based on $\alpha_{t}$ and $\bm{\beta}_{t}$ as follows:
\[
 \textbf{Influence}(s,t) = \alpha_{t}\textbf{W}_{out}(\bm{\beta}_{t} \odot \textbf{W}_{emb}[:,s]) 
\]
where $W_{emb}$ is the weight matrix of the input layer that transforms the input sequence into feature vectors, and $W_{out}$ is the weight matrix of the output layer (i.e., the softmax-layer) that transforms the latent vector into probabilities.

\begin{figure}[!htb]
    \centering
    \includegraphics[width=\columnwidth]{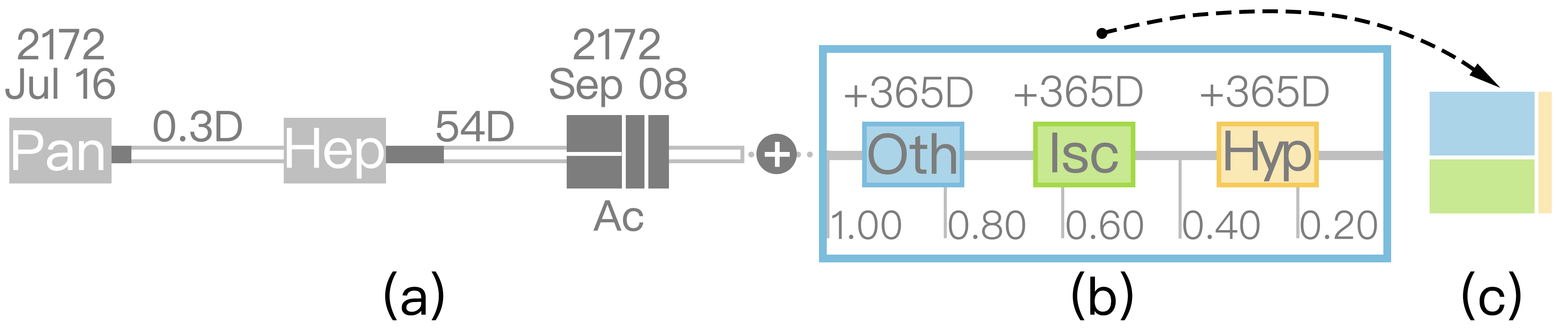}
    \caption{The visual design of the prediction view includes (a) the medical event sequence leading up to the time point of prediction, and (b) the prediction box showing the most likely diagnoses in order of predicted probabilities.}
    \label{fig:prediction} 
\end{figure}

\subsubsection{Visualization} We represent a patient's electronic medical record as a sequence of medical events, which are displayed using rectangular nodes arranged horizontally in order of event occurrence as shown in Fig.~\ref{fig:prediction}(a). To avoid overlaps (during periods of time with multiple medical events) and large gaps (during periods of time where medical events are infrequent), the event nodes are spaced with equal distance between them. The actual event times are marked above the event nodes using text labels. 

Successive event nodes are depicted with a duration bar connecting the nodes, and each bar is labeled with the time span between events.  When multiple events occur at the same time (as is common in medical data), a treemap-based representation is used to compactly represent the multi-event information within a single rectangular node.  All events are color-coded by event type, with dark gray representing treatments and light gray representing diagnoses. Hovering the mouse over on event node causes the node to be highlighted in orange and triggers the display of a tooltip showing additional event details.
Scrolling and zooming operations allow for further exploration of the patient's medical history.

The prediction results are visualized within a box located to the right of the event sequence visualization (Fig.~\ref{fig:prediction}(b)). The prediction box contains a series of rectangular nodes, one for each of the most likely predicted diagnosis events for the patient.  Each rectangular nodes is color-coded by diagnosis type, where the set of possible diagnoses are pre-chosen by a physician using the dropdown list shown in Fig.~\ref{fig:interface}(b-1). This choice is determined by the physician based on their pre-diagnosis of the patient's condition. 

The order (from left to right) of the event nodes inside the prediction box are determined by the predicted occurrence probability of the events. Therefore, the left-most event box within the prediction box corresponds to the diagnosis that is predicted to be most likely for the patient.  The predicted likelihood of each diagnosis event decreases as the boxes move toward the right of the prediction box.  The color saturation for each box indicates the prevalence of the predicted diagnosis within the medical records for a population of similar patients.

Users can clicking on a diagnosis event to view more details about the predicted diagnosis.  Available information includes a general description of the diagnosis, symptoms, causes, diagnosis methods, treatments, and typical prognosis.  These details are displayed within the description view for physicians to review.

\begin{figure}[!htb]
    \centering
    \includegraphics[width=\linewidth]{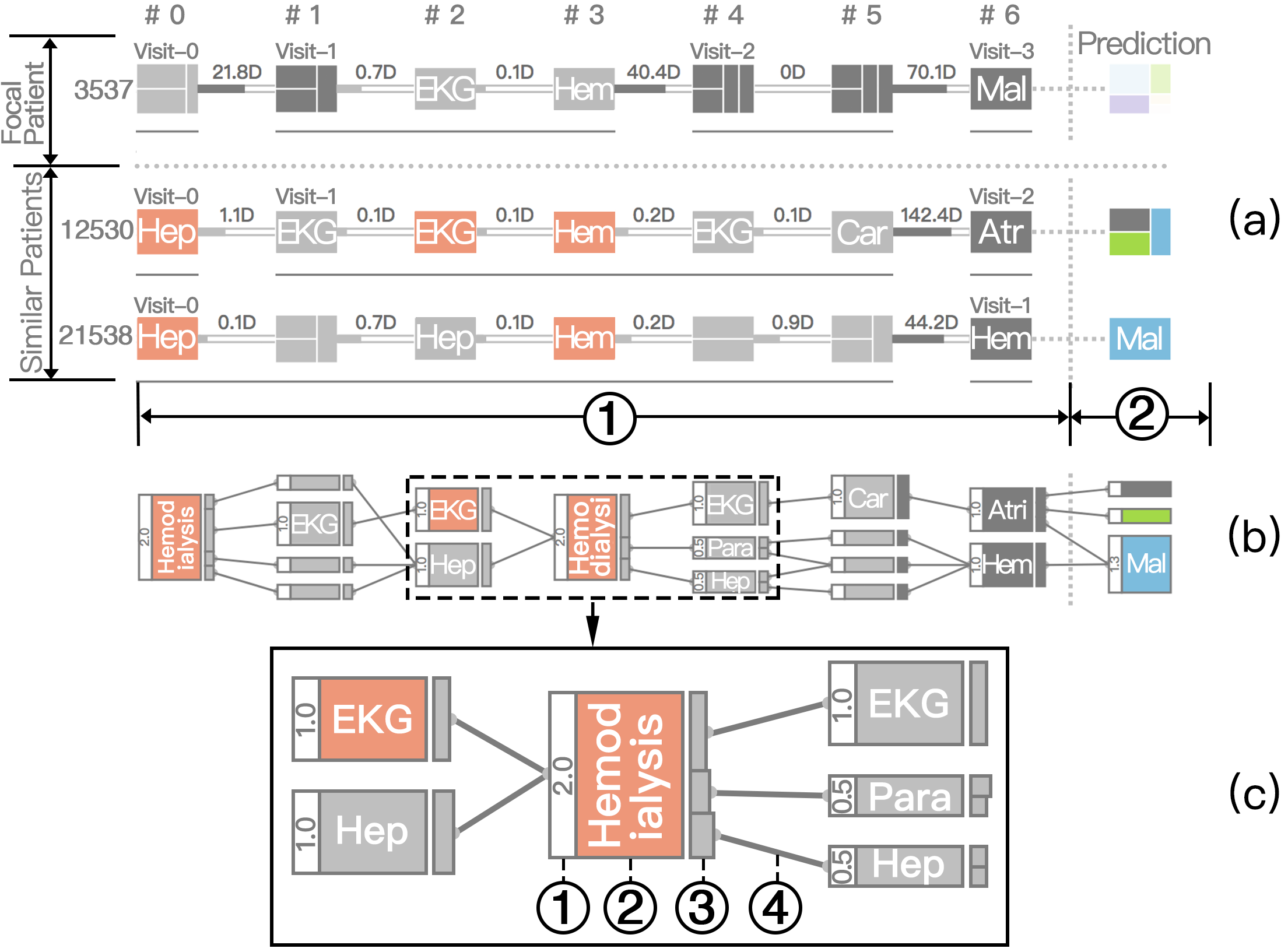}
    \caption{Medical event sequences for similar patients are visualized as either (a) individual sequences or (b) an aggregated flow diagram. Part (c) shows a more detailed illustration of the aggregate view.}
    \label{fig:similar} 
    \vspace{-0.4cm}
\end{figure}


\subsection{Similar Patient Retrieval and Comparison}
\label{sec:similar}
As identified in the pilot study, a key requirement for clinicians is the ability to compare the focal patient and prediction results to other patients with similar medical records (\textbf{R3}). \name allows users to retrieve similar patients in two ways: (1) via brushing a patient similarity histogram (Fig.~\ref{fig:interface}(d)), and (2) via explicit queries using key medical events (Fig.~\ref{fig:interface}(e)). Similar patients retrieved via either interface are displayed in a patient list (Fig.~\ref{fig:interface}(f)) which depicts a detailed event sequence for each similar patient (Fig.~\ref{fig:interface}(g)) to allow detailed comparisons.

\subsubsection{Patient Similarity and Sequence Alignment.} To support the above functions, \name adopts a distance measure to quantify the similarity between events sequences that is robust to differences in sequence length and timing. To this end, \name uses the event-to-vector and sequence alignment techniques first introduced in ET$^2$~\cite{guo2018visual}. Specifically, a vector representation of each event in a set of sequences is first calculated based on a neural network model. Sequences are then aligned temporally using a dynamic time working algorithm (DTW)~\cite{muller2007dynamic}, and distances are calculated using the event vectors. The algorithm measures similarity between sequences by estimating the similarity between each pair of events respectively in these sequences based on the Euclidean distance of the corresponding event vectors.


\subsubsection{Visualization.} The patient similarity view displays event sequence data for both the focal patient and the patients most similar to him/her.  The event sequences for similar patients are aligned to the focal patient and visualized in parallel as shown in Fig.~\ref{fig:similar}(a). We divide each of the similar sequences into two parts: (1) a history section, which best matches with focal patient's historical medical records up to the current point in time (Fig.~\ref{fig:similar}(a-1)), and (2) an outcome section which depicts the outcomes observed for the similar patients in comparison to the predicted outcome results for the focal patient (Fig.~\ref{fig:similar}(a-2)). This view adopts a visual design that is similar to the prediction view.   

To support more effective one-to-many comparison between the focal patient and the set of selected similar patients, we aggregate the medical event sequences for the similar patients into a flow-based visualization that illustrates the overall evolution of diseases and treatments within the group over time.
In this view, each medical event is visualized as a compound rectangular.  The height of the node represents the number of patients with the corresponding event at the corresponding time stage, which is also displayed as a text label in the leading rectangle (Fig.~\ref{fig:similar}(b-1)).  The node's middle rectangle
(Fig.~\ref{fig:similar}(b-2)) shows the event name.  Finally, several connection glyphs on the right edge of the node (Fig.~\ref{fig:similar}(b-3)) depict connections (via the linking lines) to subsequent nodes which occur in the next time stage (Fig.~\ref{fig:similar}(b-4)). The height of each connection glyph indicates the number of patients whose medical record includes the corresponding event transition.  The width of the connection glyphs corresponds to the average duration of the transition. 

\subsection{Treatment Outcome Analysis} 

The \name system provides a set of interactive analysis capabilities to identify key factors that effect the prediction results, and make more informed treatment plan decisions.   This is accomplished through interactions that edit the focal patient's event within the prediction view (Fig.~\ref{fig:interface}(b)) and visual comparison of the edited seqeunces in the outcome analysis view (Fig.~\ref{fig:interface}(h)). 

\begin{figure}[tb]
    \centering
    \includegraphics[width=\columnwidth]{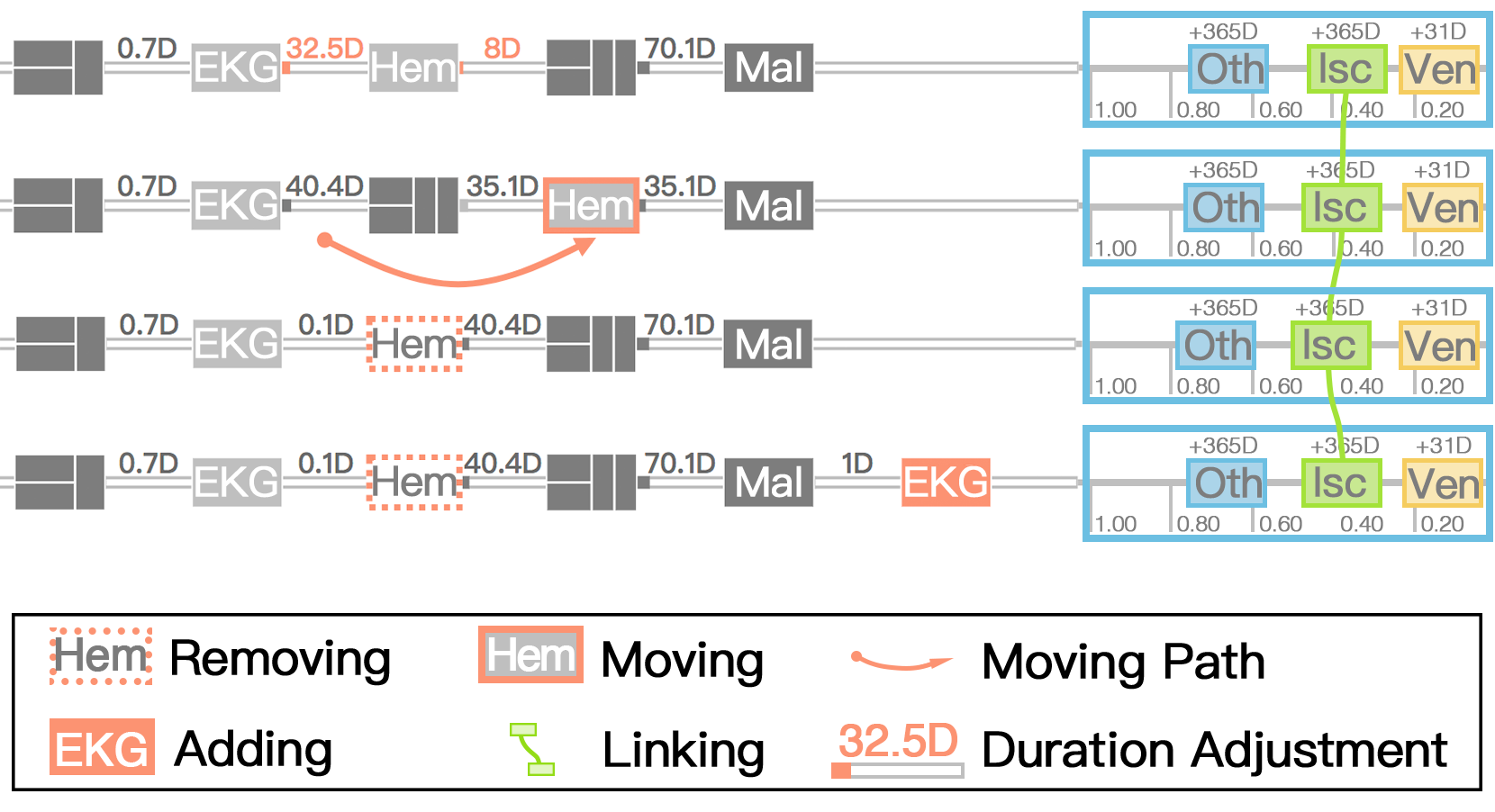}
    \caption{We enable four interactions for outcome analysis: Removing, Moving, Duration Adjustment, and Adding. The adjusted event sequences are highlighted on the left side with annotations as shown in the figure. The corresponding predicted outcomes are shown on the right side of the view.}
    \vspace{-0.1cm}
    \label{fig:outcome}
    \vspace{-0.4cm}
\end{figure}

The outcome analysis capability is summarized in Fig.~\ref{fig:outcome}.
Users can edit the focal patient's original event sequence using four interactive operations: (1) adding a new event, (2) removing an existing event, (3) adjusting the order of events, and (4) changing the duration between events.
Updated prediction results are calculated in real time in response to any edit operation is performed, and users have the option to save an edited event sequence (and resulting prediction) as a new entry within the outcome analysis view.  This allows clinicians to compare edited event sequences to explore how changes in a patient's medical record (i.e., a new treatment, or the absence of a co-morbidity) impacts the prediction results.  To support this activity, the view highlights each of the user's event sequence edits in orange (Fig.~\ref{fig:outcome}), and uses coordinated highlighting to link predictions of the same medical event across edited sequences (e.g., the same predicted diagnosis appearing for two different edited versions of the focal patient's medical record). Users can also zoom in/out on the prediction box to retrieve more detailed views.  These interactions help communicate changes in risk between sequences, especially when the same events (but with different probability) are predicted for alternative edited sequences.

A common use case for these features is when a physician investigates the potential outcomes of different treatments.  The physician can create multiple edited sequences by adding the potential treatment events to the end of the focal patient's original medical record. Viewing the predicted results under the assumption of alternative treatments can help the clinician understand the impact of each treatment option.  Alternatively, a physician could create alternative versions of a patient's medical record by removing individual events.  This would facilitate model interpretation by allowing a clinician to see the impact of a given feature on the prediction result.

Finally, to support further analysis of the contribution of key events to the predicted outcomes, \name computes the degree to which of each potential treatment is associated with each of the prediction targets within the similar patient population.  These associations are displayed in the significance view (Fig.~\ref{fig:interface}(i)) as a matrix where each row is a treatment group and each column is a predicted disease. The rows are clustered to group related treatments using the event-vector technique presented earlier in this section. 

Each cell in the matrix includes a diagram that shows the change of a disease's mean occurrence probability (shown in y-axis) and 95\% confidence interval within the subset of similar patients with the treatment (left plot) vs. those without the treatment (right plot). Cells with statistically significant differences are highlighted with a white background.    

\section{Evaluation}
\label{sec:eva}

This section presents results from three forms of evaluation: quantitative experiments to measure the prediction model's performance, a case study with a senior Chinese physician, and interviews with three physicians in both USA and China. 

\begin{table}[tb]
    \centering
    \begin{tabular}{c|c|c}
        \hline
         & Retain & Retain Extended \\ \hline
         Neg Log Likelihood & $0.2834 \pm 0.0036$ & $\textbf{0.2768} \pm 0.0036$ \\ \hline
         AUC & $0.8294 \pm 0.0022$ & $\textbf{0.8307} \pm 0.0026$ \\ \hline
         Precision & $0.8126 \pm 0.0053$ & $ 0.8126 \pm 0.0054$ \\ \hline
         Recall@2 & $0.6859 \pm 0.0081$ & $\textbf{0.6943} \pm 0.0082$ \\ \hline
         Recall@4 & $0.8954 \pm 0.0027$ & $\textbf{0.8973} \pm 0.0032$ \\
         \hline
    \end{tabular}
    \caption{Comparison of prediction performance}
    \label{tab:performance}
    \vspace{-1cm}
\end{table}

\begin{figure*}[!tb]
\centering
    \includegraphics[width=\linewidth]{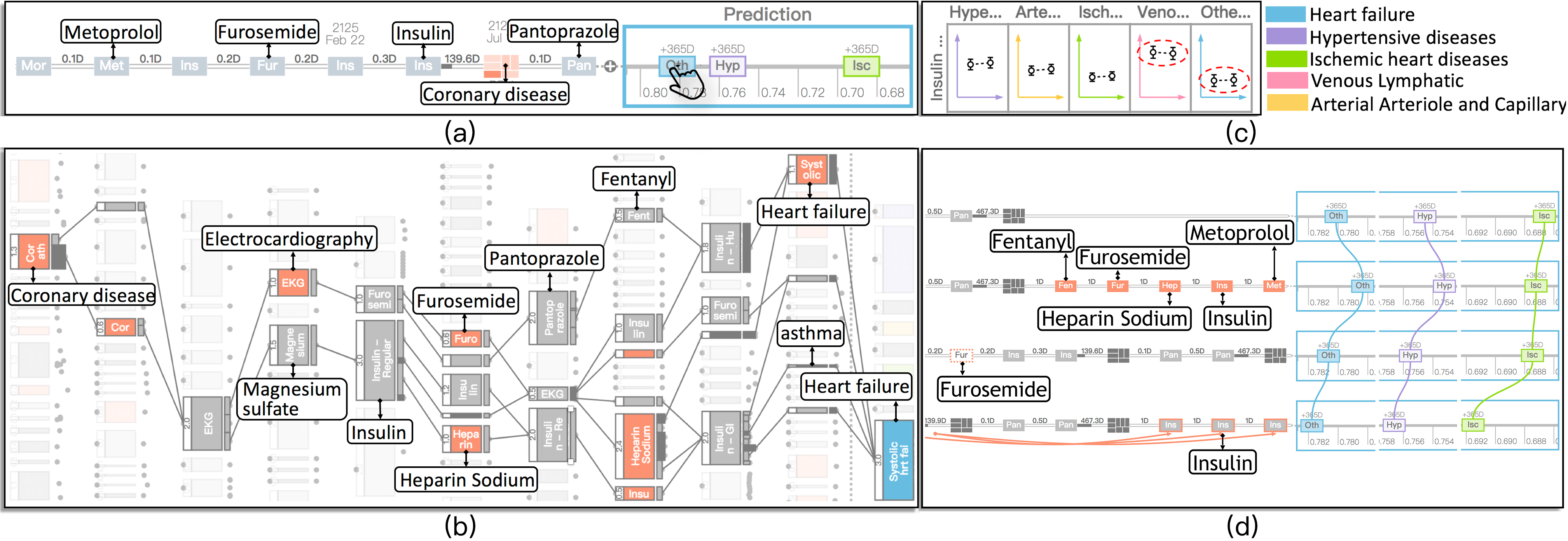}
    \caption{A case study based on a subset of MIMIC data. The results shown in this figure were identified by our expert user.}
    \vspace{-0.4cm}
    \label{fig:heart}
\end{figure*}

\subsection{Evaluation of the Prediction Model}

We compared the performance of our model (with extensions for multi-event prediction) to the original Retain single-event prediction model~\cite{choi2016retain} from which our model was derived. More specifically, medical records for patients with cardiovascular disease and at least one hospitalization were chosen from the MIMIC dataset, and their medical records were transformed into one or more event sequences based on a 6-month time window. Each sequence ended at a hospital admission event and started six months prior. As a result, 7,537 patients were selected and 64,269 sequence samples were generated.  These samples were divided into the training and testing sets using a 7:3 ratio. We further cleaned the sequences by preserving only diagnosis and treatment events. Both the original and extended Retain models were trained using the training samples to predict the risk of five highly prevalent heart and cardiovascular diseases. The disease risks were simultaneously predicted using a single extended Retain model. Meanwhile, five independent models (one for each disease) were trained for the original Retain model design.  The performance measurements in Table~\ref{tab:performance} show that our extended model performed similarly to (slightly better than) the original.

\subsection{Case Study}
We conducted a case study with a senior inpatient cardiovascular doctor with over 20 years' clinical experience in China. During the study, we first introduced \name system and the doctor was invited to use the system for himself. After getting familiar with the system's functions, the doctor were asked to perform a series of tasks including interpreting prediction results, making a treatment decision, and estimating the future outcomes for different treatment plans. The study lasted for about two hours, and the doctor was encouraged to ask questions or make comments at any time.

Fig.~\ref{fig:heart} shows the results of our study. After reviewing the patient's historical medical records (Fig.~\ref{fig:heart}(a)), the doctor said ``this patient is being treated with some typical medicines such as metoprolol and furosemide.'' He also noted that the patient suffered from diabetes after noticing regular insulin injections within their medical records. According to the prediction results, the patient had a high risk of heart failure in the future. The doctor mentioned ``it is possible as the diabetes may lead to coronary disease and finally develop into heart failure.'' The doctor was impressed by the similar patient view (the aggregated form) as shown in Fig.~\ref{fig:heart}(b). He selected the group of patients with heart failure for further inspection, and the corresponding disease progression paths were automatically highlighted by the system. He believed this view was ``very informative,'' and that the click-to-highlight function was able to ``clearly show the progression of an outcome in context of treatments.'' He felt this view gave him more confidence in the prediction results as ``it provides specific evidence [to support the prediction results].'' He also mentioned that this view would be particularly useful for medical researchers as ``it illustrates many examples following different treatment plans'' (see annotations in Fig.~\ref{fig:heart}(b)).

The doctor was also interested in the system's outcome analysis function. Specifically, he first made a care plan for the focal patient by adding multiple treatments (e.g., fentanyl, furosemide, insulin, and metoprolol) to the end of the patients existing medical record as shown in Fig~\ref{fig:heart}(d). In response, the risk of heart failure and hypertensive disease both decreased. Next, the doctor removed all of the events for furosemide (a common medicine used for heart failure patients to treat fluid build-up) from the sequence, resulting in a significant increase heart failure risk.  This also revealed in the view seen in Fig.~\ref{fig:heart}(c), showing, for example, that injections of insulin had a significant effect on reducing the risk of heart failure. Finally, the doctor moved all insulin to the end of the sequence to mimic a scenario in which the patient delayed diabetes treatment. This resulted in a further increase in heart failure risk.  The correctness of the various predictions were verified by the doctor.

\subsection{Domain Expert Interview}
In addition to the case study, we conducted in-depth interviews with three other senior physicians. To avoid bias, these doctors were different from the doctors interviewed in the pilot study. Two of the three physicians were Chinese doctors (\textbf{$E_{1,2}$}), each with 15 years of clinical experience in treating cardiovascular diseases. The third expert (\textbf{$E_{3}$}) was a senior physician in USA. During each interview, doctors were given a half-hour interactive demonstration of our system. The doctors were then invited to use the system on their own. We provided the doctors with the case study tasks as a reference, but they were encouraged to freely experiment with the system. After the experts were finished exploring the system's functionality, we conducted a semi-structured interview which incorporated several questions about overall usefulness, ease of use, general pros and cons of the prototype system, visualization designs, and insights obtained from using the system. Each interview lasted for approximately 1.5 hours, and the entire session was recorded. We summarize the collected feedback into three themes.

\subsubsection{Diagnosis Support}
All three experts appreciated our system design and believed that estimating the risk of potential diseases for a patient was useful and helpful in their daily work. All three felt that the system could contribute to an improved diagnosis process. $E_{1}$ said ``I need to take care of over 50 patients a day, ..., sometimes I am just too tired to avoid making mistakes, ..., if the system is developed based on statistical analysis of similar medical records, I'd love to trust the results, ..., and it can actually help us reduce the risk of making a mistake.'' $E_2$ believed that \name would particularly useful for inexperienced doctors or medical students in helping them make more accurate diagnoses.  $E_2$ also mentioned that ``this tool can [help] reduce a doctor's burden.''  Similarly, $E_{3}$ mentioned that ``doctors' time is valuable, quickly estimating the risk of a patient [using the system] is a useful function.''

\subsubsection{Similar Patient Retrieval and Comparison} 
All three experts felt that the similar patient view (as well as the similar patient retrieval mechanisms) were useful. $E_1$ said that ``comparing to the similar patients in detail not only gives me more confidence of the prediction results but also provides me with rich treatment examples.'' $E_2$ said that the ``medical records of similar patients are an important reference for a doctor to make a proper diagnosis, but sometimes the doctor cannot fully review a patient's entire medical records [due to limited time or unavailable of the data]..., the system provides a more efficient way for us to retrieve the similar patients [when compared to the system we are currently using].'' $E_3$ felt that ``the most valuable part of the system would be the impact different treatment approaches would have on similar patients.'' They also believed that the similar patient view, especially the aggregated representation, was ``complex and took time to learn.'' However, once they got familiar with the design, they felt this view was ``informative'' and ``clearly illustrated different care plans and the corresponding outcomes.''

\subsubsection{Treatment Outcome Analysis}
Both $E_{1,2}$ believed that the treatment outcome analysis feature in \name system was a highlight. $E_1$ believed that the idea of virtually making different care plans and comparing their potential outcomes was a ``cool and valuable'' feature to support making a prognosis. However, he also mentioned that the usefulness relied heavily on the precision of the prediction results. $E_2$ believed that ``this system provided an interactive way for exploring some complicated situations and their influence on the patient.'' However, she also reminded us that a prognosis estimate is usually based on the statistics of a very large collection of patients over a very long period of time.  She pointed out that this feature, therefore, was useful only when the underlying data was rich enough to represent the rich variety of outcomes that patients face. $E_3$ believed that doctors only cared about the end results. Thus, if the system could directly provide the factors that influences the outcome, it will be more efficient.

\begin{figure}[tb]
    \centering
    \includegraphics[width=\columnwidth]{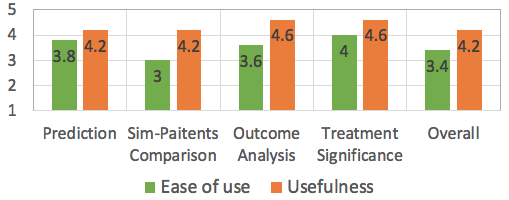}
    \vspace{-0.6cm}
    \caption{The questionnaire results}
    \label{fig:questionnaire} 
    \vspace{-0.3cm}
\end{figure}

\subsection{Post-Interview Questionnaire} All four of the experts (3 interviewees, and 1 from the case study) were invited to complete a questionnaire after their interview.  The questionnaire asked doctors to rate the ease-of-use and usefulness of the key features of our system on a scale of 1-5, with 1 indicating difficult to use / useless and 5 indicating very easy to use / very useful. The results are summarized in Fig.~\ref{fig:questionnaire}. The primary issue highlighted in these results were related to ease-of-use, which we discuss in the next section.

\section{Discussion}
\label{sec:diss}
The results from our case study and expert interviews were generally positive, with users confirming the usefulness of \name and expressing excitement regarding the treatment outcome analysis capabilities. However, they also identified several limitations, provided a number of constructive suggestions, and mentioned some interesting potential applications of the system. 

\subsection{Limitations} 
The major limitations of \name system include ease-of-use, data quality, and a lack input methods.

\subsubsection{Ease-Of-Use} Although all of the experts acknowledged the usefulness of the system, they also mentioned that learning how to use it took essential time away from a doctor. All of the experts were familiar with traditional statistical diagrams such as histograms and line charts, however they needed some practice to read some of the more complex views introduced in \name. However, they also believed these new designs were more informative when compared to more familiar statistical charts.  They also felt that the difficulty mainly comes from the lack training. For example, $E_2$ said ``we (doctors) spend years in school to learn how to make diagnosis based on those [traditional] statistical tools and diagrams, ..., your tool is obviously more informative but we just need more time to get familiar with it.'' Second, operating multiple coordinated views also take some effort. Both $E_{1,3}$ mentioned that it would easier to use if the tool could directly generate and print out a report without as many interactions. $E_3$ also said ``it will be easier to use if you could somehow separate the views of three different functions apart into multiple pages and guide the operation in a step to step manner instead of packing them all together.''

\subsubsection{Data Quality} The Chinese physicians were concerned about the quality of the training data which directly influenced on our analysis results. Both $E_{1,2}$ mentioned that the quality of the electronic health records collected in Chinese hospitals were much worse than that of the MIMIC dataset. They mentioned that in medical data in China was primarily free-text, and that many hospitals in China were just started to use electronic health record systems. That limits the longitudinal extent of data that could be used as input to the system. For this reason, they believed that \name might not be as useful in Chinese hospitals right away. $E_3$ also mentioned that the treatment outcome analysis should be based on a larger dataset collected within a longer time window (e.g., several years). Although the current system demonstrated the usefulness of the function, more data will need to be imported into the system before being applied in real clinical scenarios. 

\subsubsection{Lack of input methods} $E_1$ also felt that although \name was useful, the design was not sufficient as it has limited ways for clinicians to enter new medical data. In particular, she said ``when compared to the existing system, your tool focus more on the analysis, but lacks of a convenient method for me to enter medical records in the text form''.


\subsection{Implications}
Our experts also raised many implications of the system, which can be summarized into two broad categories.

\subsubsection{From Knowledge-Sharing to Experience-Sharing} All of the experts believed that \name would be especially useful for junior physicians, medical students or other inexperienced health professionals. They believed that since the prediction model in \name is trained based on the treatment records made by experienced physicians, it would capture those doctors' experiences. In comparison, most existing knowledge-based systems only share medical knowledge. In particular, $E_{1,2}$ mentioned that in China there are many undeveloped rural areas with poor health systems where doctors are less experienced and less well-trained. The \name system would help provide information to these doctors based on the experience of more senior clinicians.  This maps to typical doctor training techniques, where doctors first learn from medical textbooks before a long period of training under the supervision of senior doctors to help them gain knowledge through experience.  


\subsubsection{From Doctors to Other Users} Our experts also suggested many other potential application scenarios for the \name system. For example, $E_1$ believed that our system would be very useful for analysts in a medical insurance company. ``It can help an insurance company estimate the risk of a patient in a more efficient way" said by $E_1$. Both $E_1$ and $E_2$ mentioned that our tool could be very helpful for medical research as it is ``build based on statistical analysis and provides many advanced visual diagrams, illustrating the evidence of the analysis results". Both $E_2$ and $E_3$ felt that \name system could also be directly used by a patient as ``it suggests the risk a patient may have" and  ``the patient may want to spend more time investigating the functionality of the system". These scenarios greatly expend the application scope of \name system, though certain design changes may be required for different applications.

\section{Conclusion}
\label{sec:conlusion}
This paper introduced an intelligent clinical decision assistance system, \name, that uses large-scale EHR data to help physicians make decisions during their clinical workflow. The system, designed based on requirements identified in a pilot study, provides clinical assistance through a state-of-the-art deep learning prediction model as well as an interactive visual interface for exploration and interpretation. The interaction pipeline of our system, consists of three major steps: (1) diagnosis support, (2) similar patients retrieval and comparison, and (3) treatment outcome analysis. 
We evaluated the system via a case study, expert interviews, and a quantitative evaluation of the predictive model.
The results from these evaluations showed that the overall system provided valuable assistance to the clinical decision process.  In the future, we plan to address the aforementioned issues and conduct a larger evaluation of the system in a local hospital so as to update the system's models based on local patients' conditions.      


\bibliographystyle{IEEEtran}
\bibliography{ms}

\end{document}